\documentclass{ocephys}

\DeclareSIUnit{\days}{days}
\DeclareSIUnit{\day}{day}
\newcommand{\Ro}{\mathrm{Ro}}

\renewcommand{\vec}[1]{\boldsymbol{#1}}

\renewcommand{\d}{\mathrm{d}}
\newcommand{\pp}[2]{\frac{\partial #1}{\partial #2}}

\DeclareMathOperator{\J}{J}

\title{Global Near-Inertial Wave Spectra Shaped by Mesoscale Eddies}

\author{Scott Conn}
\author{J\"orn Callies}
\affil{California Institute of Technology, Pasadena, California}
\corr{Scott Conn, \href{mailto:sconn@caltech.edu}{sconn@caltech.edu}}

\begin{document}
\nolinenumbers

\section*{Abstract}

Wind-forced near-inertial waves (NIWs) propagate through a sea of mesoscale eddies, which can fundamentally alter their evolution. The nature of this NIW--mesoscale interaction depends on how dispersive the waves are. For weakly dispersive waves, ray tracing suggests that the NIW frequency should be shifted by $\frac{1}{2}\zeta$, where $\zeta$ is the mesoscale vorticity, and that the waves are refracted into anticyclones. Strongly dispersive waves, in contrast, retain the large-scale structure of the wind forcing and exhibit a small negative frequency shift. Previous in situ observational studies have indeed revealed varying degrees of NIW--mesoscale interaction. Here, observations of NIWs from drifters are used to map the geography of NIW--mesoscale interactions globally, and idealized simulations and a simple model are used to identify the underlying physical processes. Almost everywhere in the ocean, with the notable exception of the North Pacific, the NIW frequency is strongly modulated by the mesoscale vorticity, with the slope of the frequency shift vs.\ vorticity taking values of approximately~$0.4$. Concentration of NIW energy into anticyclones is a common feature throughout the ocean. Other aspects of the observations, however, show signatures of strongly dispersive waves: a negative frequency shift and weaker concentration into anticyclones in regions with strong eddies as well as weak modulation of the NIW frequency by mesoscale eddies in the North Pacific. The signatures of both weakly and strongly dispersive NIW behavior can be rationalized by the geography of the wave dispersiveness and the fact that wind forcing excites multiple vertical modes with different wave dispersiveness. These results have implications for NIW-induced mixing in the upper-ocean.

\clearpage

\section{Introduction}

Near-inertial waves (NIWs) excited by atmospheric storms substantially influence upper-ocean dynamics through their role in small-scale mixing \parencite[e.g.,][]{alford2016}. They are characterized by strong vertical shear that can trigger shear instabilities, causing turbulence and mixing. This process deepens the surface mixed-layer in the aftermath of a storm, and a significant fraction of NIW energy is dissipated in the upper ocean \parencite{alford2001,plueddemann2006observations,alford2020}. Ultimately, NIW-induced mixing influences the ocean’s surface heat budget, impacting atmospheric circulation and precipitation patterns \parencite{jochum2013}.

The evolution of NIWs can be strongly influenced by interactions with mesoscale eddies. If the waves are weakly dispersive, they can be described by ray tracing \parencite{kunze1985, conn2025regimes}, and one expects them to be refracted toward anticyclonic vorticity. There is now substantial observational evidence for such $\zeta$~refraction and the resulting concentration of NIW energy into anticyclones \parencite{elipot2010modification, thomas2020, yu2022, conn2024interpreting}. This refraction also increases the speed at which NIWs propagate to depth \parencite{lee_inertial_1998, asselin2020penetration}, which in turn impacts where in the upper ocean NIWs generate mixing \parencite[e.g.,][]{essink2022, alford_observations_2025}.

The recent evidence of $\zeta$~refraction from field campaigns in the North Atlantic \parencite{thomas2020, yu2022, conn2024interpreting} stands in contrast to the pioneering Ocean Storms Experiment in the Northeast Pacific \parencite{dasaro1995a}, where the NIW--mesoscale interaction was found to be weak. Studying the scale reduction of the NIW field after the passage of a storm, \textcite{dasaro1995a} could explain the evolution as driven by the $\beta$~effect, so-called $\beta$~refraction \parencite{munk_coherence_1968,gill1984}, with mesoscale eddies playing no discernible role \parencite{dasaro1995c}. \textcite[][hereafter YBJ]{young1997} argued that the lack of NIW--mesoscale interaction during Ocean Storms was due to a breakdown of the scale separation assumption of ray tracing, showing that strongly dispersive waves remain largely uniform in the presence of eddies. Indeed, \textcite{thomas2024near} showed that the differences in NIW behavior between Ocean Storms and the NISKINe experiment in the North Atlantic could be attributed to differences in how dispersive the waves were in these two regions.

Given these drastic differences in NIW behavior, the goal of this paper is to characterize the geography of NIW--mesoscale interaction globally. We use YBJ's description of the NIW evolution in the presence of a mesoscale eddy field both as a guide in the analysis and as our main interpretive framework. Throughout the paper, we consider the YBJ equation for a single vertical mode, which requires assuming a barotropic eddy field. For a vertical mode with structure $g(z)$, the NIW velocity can be expressed as $(u_w(x,y,t), v_w(x,y,t))g(z)$. The YBJ equation is then formulated as an evolution equation for $\phi=(u_w + i v_w)e^{i f t}$, where $f$ is the inertial frequency and here assumed constant (i.e., we neglect the $\beta$~effect). Under these assumptions, the YBJ equation can be written as 
\begin{equation} 
  \pp{\phi}{t} + \J(\psi,\phi) + \frac{i\zeta}{2}\phi - \frac{i f \lambda^2}{2}\nabla^2\phi = 0,
  \label{eq:YBJ}
\end{equation} 
where $\psi$ is the prescribed streamfunction of mesoscale eddies, $\zeta = \nabla^2 \psi$ is the associated vorticity, $\lambda$ is the deformation radius of the vertical mode under consideration, and $\J(a,b) = \partial_x a \, \partial_y b - \partial_y a \, \partial_x b$ is the Jacobian operator. The second term in~\eqref{eq:YBJ} represents the advection of NIWs by mesoscale eddies, the third term represents refraction by mesoscale vorticity, and the fourth term represents dispersion. The relative importance of dispersion versus refraction is captured by the wave dispersiveness $\varepsilon^2 = f \lambda^2 / \Psi$, where $\Psi$ is an appropriate scale for the mesoscale streamfunction.

\begin{figure}
  \centering
  \includegraphics{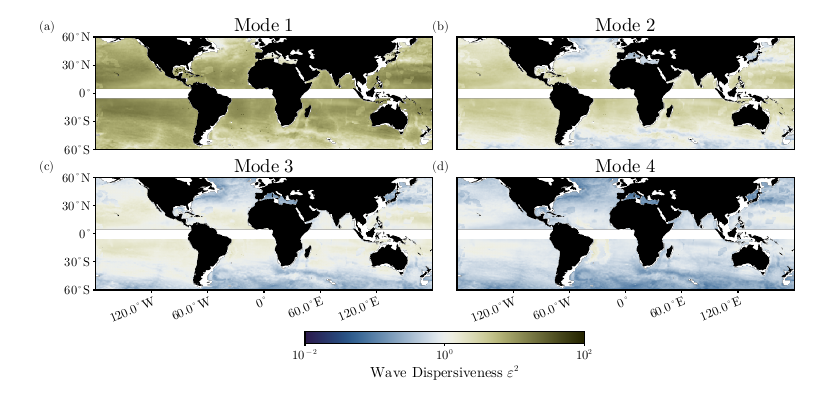}
  \caption{Geography of the wave dispersiveness~$\varepsilon^2$ of the first four vertical modes, showing a transition from strongly dispersive low modes to more weakly dispersive high modes. Note the weaker dispersion in major current systems. Figure adapted from \textcite{conn2025regimes}.}
  \label{fig:eps}
\end{figure}

The wave dispersiveness $\varepsilon^2$ governs the extent to which NIW--mesoscale interactions influence the evolution of the waves (Fig.~\ref{fig:eps}). When $\varepsilon\gg 1$, dispersion dominates and mesoscale effects are limited. When $\varepsilon\ll 1$, dispersion is weak and the mesoscale strongly modulates NIW behavior. \textcite{kunze1985} used ray-tracing to describe how NIWs interact with mesoscale eddies, proposing that the eddies act to shift the local inertial frequency to an effective value $f + \zeta/2$. This ray-tracing framework applies in the weak-dispersion limit, $\varepsilon \ll 1$, when the scale separation assumption is appropriate, even when the atmospheric forcing is large-scale \parencite{conn2025regimes}, providing a useful prediction for how mesoscale eddies should modulate the NIW frequency. This frequency shift also sets up phase gradients in the NIWs. Dispersion acts to flux energy down these phase gradients, resulting in the concentration of NIW energy into anticyclones, although advective straining can complicate this process \parencite{rocha2018}. If dispersion is strong, in contrast, the waves only take on weak horizontal structure that is proportional to the streamfunction~$\psi$ and suppressed by a factor of order~$\varepsilon^{-2}$ compared to the leading-order uniform structure \parencite{young1997,conn2025regimes}. Similarly, the NIW frequency shift in this regime is suppressed by a factor of order~$\varepsilon^{-2}$ compared to the scale of~$\zeta$ and proportional to the area-averaged mesoscale kinetic energy rather than~$\zeta$ \parencite{young1997,conn2025regimes}:
\begin{equation}
    \Delta\omega = -\frac{1}{f\lambda^2}\frac{\int \frac{1}{2}|\nabla\psi|^2 \, \d^2 x}{\int \d^2 x}.
    \label{eq:sd_freq_shift}
\end{equation}
\textcite{conn2025regimes} showed that a horizontally uniform wind forcing primarily projects onto a single mode, meaning the spectrum of the NIW should be strongly peaked in this strong-dispersion case.

The global availability of surface drifters and satellite altimetry provides an opportunity to assess the importance of NIW--mesoscale interactions across the world ocean. In~situ field campaigns offer detailed spatial information over limited regions but cannot achieve global coverage. Altimetry by itself is blind to NIWs, as they produce no leading-order pressure signal, but it can be used to characterize the mesoscale eddy field. Drifters often exhibit easily discernible inertial circles \parencite[e.g.,][]{dasaro1995a,beron-vera_statistics_2016}, and frequency spectra universally exhibit a distinct near-inertial peak \parencite{yu_surface_2019}. The drifter observations' Lagrangian nature means that they capture the waves' intrinsic frequencies and are not affected by Doppler shifts.

\textcite{elipot2010modification} previously characterized the global spectral properties of NIWs using drifter data. Averaged globally, they found that the frequency of the NIW peak roughly followed with $0.4\zeta$, close to the $\zeta/2$ prediction from \textcite{kunze1985}. They identified wind forcing as the primary driver of NIW amplitude variations but also found $\zeta$ to modulate the amplitude, consistent with ray tracing and YBJ theory. \textcite{elipot2010modification} further linked variations in the width of the NIW peak to the Laplacian of~$\zeta$, based on theoretical arguments by \textcite{klein2004organization}.

The strength of mesoscale eddies, the inertial frequency, the deformation radius, and the projection of wind forcing onto vertical modes all vary substantially across the world ocean. Because these quantities shape the characteristics of NIW--mesoscale interactions, one should expect substantial regional differences as exemplified by the dichotomy between Ocean Storms and NISKINe \parencite{thomas2024near}. These regional differences can be obscured in a global mean, so we leverage the expanded drifter dataset available since \textcite{elipot2010modification} to explore this geography.

This analysis shows that mesoscale eddies modulate the NIW frequency nearly everywhere, with the notable exception of the North Pacific. In highly energetic regions, we also find evidence for the excitation of strongly dispersive waves. We interpret these results emphasizing the role of the wave dispersiveness in modulating NIW--mesoscale interactions. Idealized simulations and a simple model of this interaction reproduce key spectral features observed in the drifter data. Together, these results show how NIW--mesoscale interactions have a major organizing influence on the global structure of near-inertial energy.

\section{Drifter Observations of NIWs}

\begin{figure}[t]
    \centering
    \includegraphics{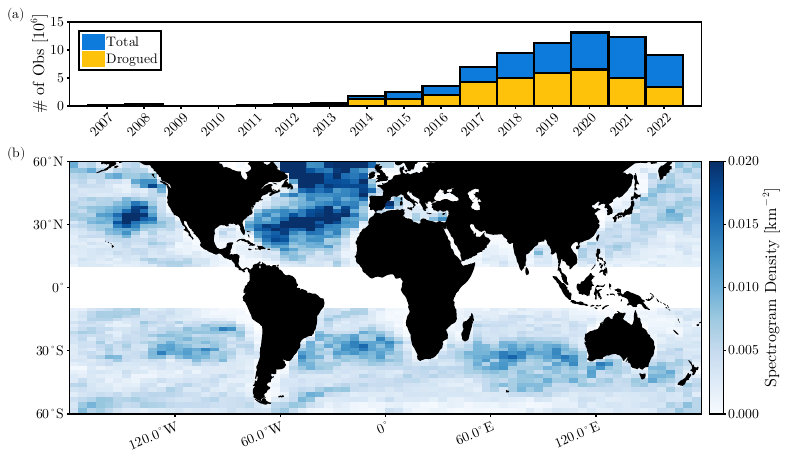}
    \caption{Availability of drifter data. (a)~Number of hourly drifter observations available per year. Blue refers to the full set of GPS-tracked drifters, while yellow represents only those observations for which the drifter's drogue is still attached. (b)~Distribution of drifter observations, shown as the number of spectrograms per unit area in a given grid cell. The pattern reflects the large-scale convergence and divergence in the ocean as well as a northern-hemisphere and Atlantic bias.}
    \label{fig:drifter_stats}
\end{figure}

To characterize NIWs globally, we use observations from the Global Drifter Program. This dataset provides hourly records of both position and horizontal velocity for \num{19396}~drifters spanning the period 2007–2023. We restrict our analysis to drifters with GPS-tracked positions. Each drifter is initially drogued, such that its velocity reflects the current at \qty{15}{\meter} depth; however, the drogue can be lost over the course of a drifter's lifetime. Observations made while the drogue is still attached account for approximately 49\% of the dataset (Fig.~\ref{fig:drifter_stats}a). We discard all measurements following drogue loss. The distribution of drifters across the ocean is non-uniform and reflects large-scale patterns of horizontal convergence and divergence as well as preferential deployment, for example in the North Atlantic (Fig.~\ref{fig:drifter_stats}b).

To calculate NIW spectra, we first divide the drifter trajectories into overlapping 20-day segments and compute the spectrogram of $u + iv$ for each segment. These spectrograms are calculated using a Lanczos window and subsequently binned based on the mean position of each segment, also determined using the same window. A defining feature of NIWs is their circular polarization. In the Northern Hemisphere (where $f > 0$), NIWs exhibit clockwise polarization, so we retain only the clockwise component of the power spectrum; in the Southern Hemisphere (where $f < 0$), we retain only the anticlockwise component. The 20-day segment results in a spectral resolution of \qty{3.6e-6}{\per\second}. Spectral estimates are calculated as averages over geographical bins. The grid spacing of the binning is \ang{5} zonally and variable in the meridional direction. The meridional grid spacing is capped at \ang{5} and decreases toward the equator to ensure that the variation in~$f$ across each grid cell remains smaller than the spectral resolution, rendering the impact of $f$~variations on the spectral estimates negligible.

\begin{figure}
    \centering
    \includegraphics{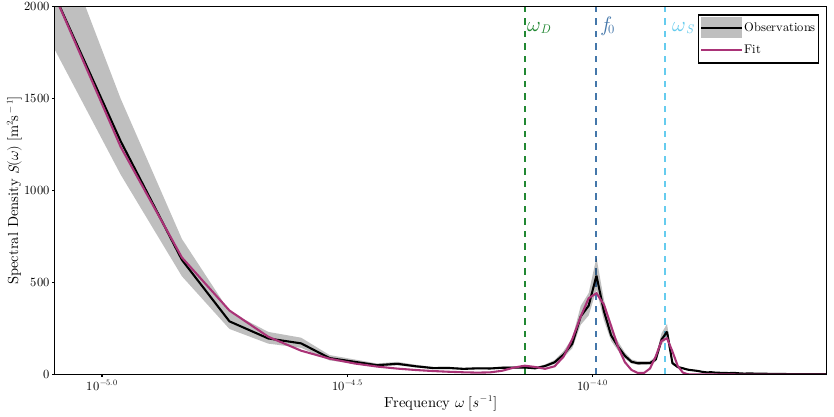}
    \caption{Example spectral estimate (black) and the associated least-squares fit of the model spectrum (purple). This example shows peaks corresponding to NIWs as well as the semidiurnal tide. There is little power associated with the diurnal tide in this example.}
    \label{fig:spec_fit}
\end{figure}

The spectra show evidence of low-frequency balanced motion, NIWs, diurnal/semi-diurnal tides, and high-frequency internal waves. We perform a spectral fit to isolate the NIW signal. The model spectrum consists of a low-frequency component and Gaussian peaks for the NIWs and tides. We perform the fit in linear space, so we do not include a term for the internal wave continuum as its amplitude is orders of magnitude smaller than the other terms. The spectral model is
\begin{equation}
    S(\omega) = \frac{A_L}{1+(\frac{\omega}{\omega_L})^s} + A_Ie^{-\frac{(\omega - f - \omega_I)^2} {2\sigma_I^2}} + A_De^{-\frac{(\omega - \omega_D)^2}{2\sigma_D^2}}+ A_Se^{-\frac{(\omega - \omega_S)^2}{2\sigma_S^2}}\label{eq:spectral_model},
\end{equation}
where $A_L,A_I,A_D,A_S$ are the amplitudes of the low-frequency motion, NIWs, diurnal tides, and semi-diurnal tides, respectively, $\omega_L$~is the transition frequency of the low-frequency model, $s$~is the spectral slope of the low-frequency model, $\omega_I$~is the NIW frequency shift, $\sigma_I$~is the NIW spectral width, $\omega_D$~is the frequency of the diurnal tides, $\sigma_D$~is the width of the diurnal tides, $\omega_S$~is the frequency of the semi-diurnal tides, and $\sigma_S$~is the width of the semi-diurnal tides. As we consider either the clockwise or anticlockwise components of the power spectrum alone, we take the frequencies to be positive. With this convention $\omega_I>0$ corresponds to a shift of the NIWs to frequencies higher than $f$ (i.e., superinertial frequencies), while $\omega_I>0$ corresponds to a shift to frequencies lower than $f$ (i.e., subinertial frequencies).

Given a spectral estimate $\hat{S}$, we fit the model spectrum to this estimate using the least squares method (Appendix~A). We fix the tidal frequencies and determine all other parameters through the fit (see Fig.~\ref{fig:spec_fit} for an example). The NIW kinetic energy can be obtained from the fit by integrating the NIW part of the model spectrum across all frequencies:
\begin{equation}
    K = \sqrt{2\pi}A_I\sigma_I.
\end{equation}
Each spectral estimate~$\hat{S}$ is obtained by averaging over all available spectrograms in the respective bin. The spectral estimate~$\hat{S}$ follows a $\chi^2$~distribution, where the number of degrees of freedom is equal to twice the number of spectrograms that are averaged to obtain~$\hat{S}$. We draw random samples from this distribution and perform the fitting procedure on each sample to obtain an empirical distribution for the fitted parameters.

\begin{figure}
  \centering
  \includegraphics{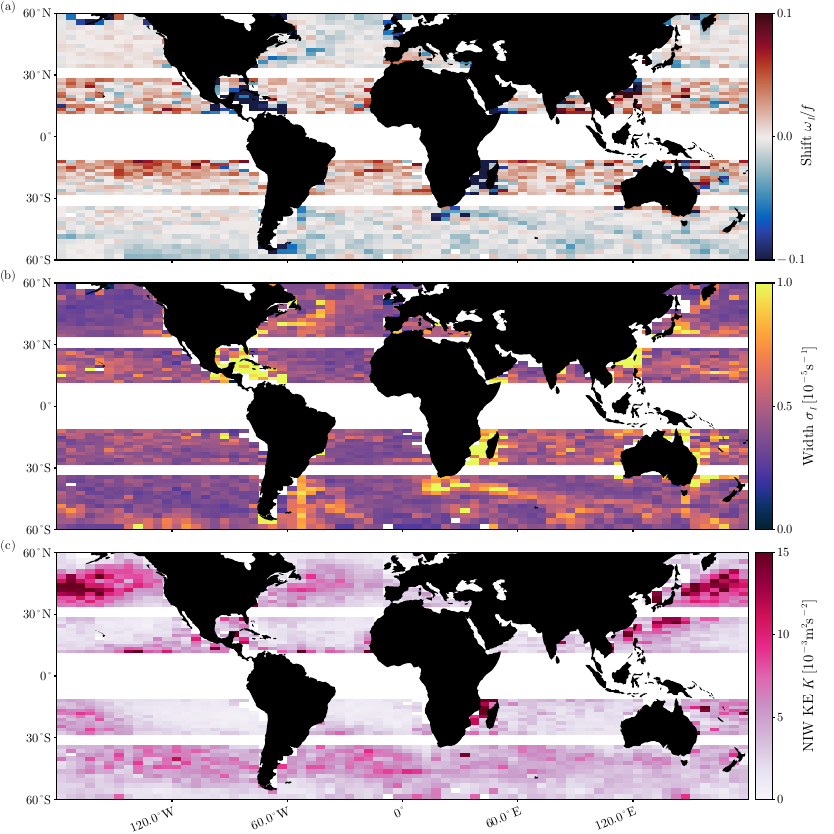}
  \caption{Characteristics of the NIW peak: (a)~the frequency shift~$\omega_I/f$, (b)~the spectral width~$\sigma_I$, and (c)~the NIW kinetic energy~$K$. All estimates are calculated from averages over all available spectrograms in a given geographical bin. 95\% confidence intervals are shown in Suppl.~Fig.~1.}
  \label{fig:global_spec_prop}
\end{figure}

First, we apply the fitting procedure to spectral estimates obtained by averaging all spectra within a given spatial bin. We exclude regions near the equator, where NIWs merge with low-frequency tropical wave modes, and around the turning latitudes of the semi-diurnal tides, where NIWs cannot be distinguished from tidal signals. Equatorward of \ang{30}, the mean frequency shift $\omega_I$ is generally positive (Fig.~\ref{fig:global_spec_prop}a). \textcite{elipot2010modification} attributed this to the equatorward propagation of NIWs—a plausible explanation that we revisit in the discussion (Section~\ref{sec:discussion}).

Poleward of \ang{30}, we observe regions with both positive and negative frequency shifts. Negative shifts occur primarily in energetic regions such as western boundary currents and the Antarctic Circumpolar Current (ACC). We hypothesize that these shifts result from the excitation of strongly dispersive NIWs, which produce a negative frequency shift as described by~\eqref{eq:sd_freq_shift}. We explore this hypothesis in more detail below. In other regions, the frequency shift is weakly positive.

One potential concern is that the observed frequency shifts may be influenced by a vorticity sampling bias inherent to the drifter data. Drifters preferentially sample regions of convergent flow, which are often associated with cyclonic structures. If weakly dispersive NIWs are present, ray-tracing predicts a positive frequency shift in such regions, potentially introducing a net positive bias. However, we will show below that the spatial patterns in the frequency shift persist even when this bias is taken into account.

The width~$\sigma_I$ of the NIW peak is remarkably uniform across most of the ocean but is elevated in western boundary currents and the ACC (Fig.~\ref{fig:global_spec_prop}b). This suggests an influence of mesoscale eddies, but strongly dispersive waves should be narrowly peaked and the ray-tracing framework for weakly dispersive waves does not provide predictions for spectral width. In the following sections, we use the YBJ model to examine the physical mechanisms responsible for the increased spectral width in high-energy regions.

The NIW kinetic energy is elevated beneath the mid-latitude storm tracks in the North Pacific, North Atlantic, and Southern Ocean, with a notable maximum in NIW kinetic energy in the North Pacific (Fig.~\ref{fig:global_spec_prop}c). This pattern broadly reflects the large-scale distribution of wind energy input into the near-inertial band \parencite{flexas2019, alford2020}. Models suggest some asymmetry in wind-work magnitude between the North Pacific and North Atlantic, although the dynamics governing NIW propagation to depth likely also influence the kinetic energy generated by a given wind forcing. We explore these dynamical differences further below.

The patterns of the NIW frequency shift~$\omega_I$ and peak width~$\sigma_I$ show structure that is spatially aligned with regions of energetic mesoscale turbulence, suggesting that interactions with mesoscale eddies influence the NIW spectral properties. To investigate this influence, we characterize the mesoscale eddy field using satellite altimetry measurements of sea surface height (SSH). We use the delayed-time (DT) 2018 release of the Data Unification and Altimeter Combination System (DUACS) product \parencite{taburet2019duacs}. 
The mesoscale streamfunction~$\psi$ is defined as $\psi = gh/f$, where $g$ is the gravitational acceleration and $h$ is the SSH. We then compute the mesoscale vorticity $\zeta = \nabla^2 \psi$ and interpolate it onto all drifter trajectories to obtain concurrent estimates of NIW velocity and mesoscale vorticity.

The influence of mesoscale eddies on NIW spectra can be assessed by evaluating NIW spectral properties as a function of $\zeta$. Within each spatial bin, we subdivide the spectrograms based on the mean vorticity along each trajectory, calculated using the same Lanczos window as above. For each vorticity bin, we compute spectral estimate and extract the spectral fit parameters defined in~\eqref{eq:spectral_model}.

\begin{figure}
    \centering
    \includegraphics{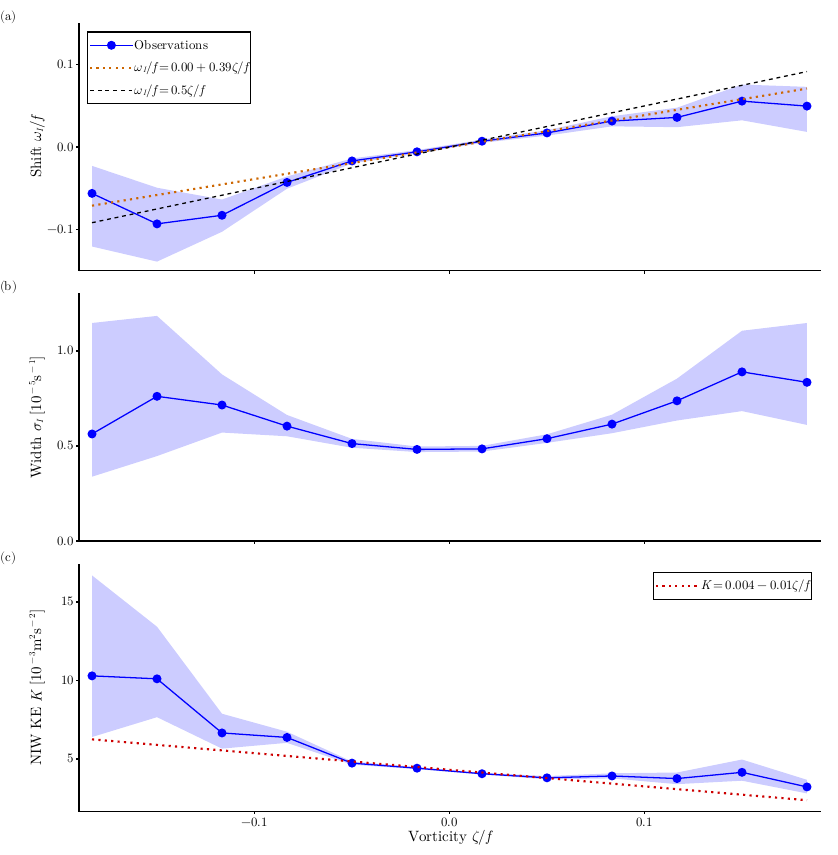}
    \caption{Globally averaged dependence of NIW spectral properties on mesoscale vorticity~$\zeta$: (a)~frequency shift~$\omega_I/f$ (blue), with the associated linear fit (orange) and the $\zeta/2$ prediction (black), (b)~spectral width~$\sigma_I$ (blue), (c)~NIW kinetic energy~$K$ (blue). Shading indicates 95\% confidence intervals.}
    \label{fig:global_av}
\end{figure}

The spectral properties of NIWs, when averaged globally, show strong evidence of modulation by mesoscale eddies (Fig.~\ref{fig:global_av}). To quantify this modulation, we perform a linear regression of $\omega_I$ and $K$ to $\zeta$:
\begin{align}
    \frac{\omega_I(\zeta)}{f} &= a + b\frac{\zeta}{f},\\
     K(\zeta) &= c + d\frac{\zeta}{f}.
\end{align}
We find that the frequency shift (Fig.~\ref{fig:global_av}a) varies with $\zeta$ with a slope of \qty{0.39} (95\% CI: \numrange{0.36}{0.42}) and an intercept of \qty{0.000} (95\% CI: \numrange{-0.001}{0.001}). This slope is in excellent agreement with the values reported by \textcite{elipot2010modification}, although our estimated intercept is lower. The dependence of the spectral width on vorticity (Fig.\ref{fig:global_av}b) also resembles the findings of \textcite{elipot2010modification}; however, large uncertainties in high-vorticity bins make it difficult to determine whether there is a robust relationship. Globally averaged, we observe a clear concentration of NIW kinetic energy into anticyclonic regions (Fig.~\ref{fig:global_av}c). The slope of the linear fit is \qty{0.0011}{\meter\squared\per\second\squared} (95\% CI: \qtyrange{0.0009}{0.0012}{\meter\squared\per\second\squared}), and the intercept is \qty{0.0043}{\meter\squared\per\second\squared} (95\% CI: \qtyrange{0.0042}{0.0044}{\meter\squared\per\second\squared}). While a linear fit is not a good model for $K(\zeta)$ at high vorticity, the slope provides a useful summary metric for quantifying the preferential concentration into anticyclones.

Caution is warranted when interpreting global averages of spectral properties. For example, the spectral width is elevated in western boundary currents and the Antarctic Circumpolar Current (ACC), where vorticity is also strong (Fig.~\ref{fig:global_spec_prop}b). The upward curvature in the width--vorticity relationship may therefore reflect a coincidental spatial correlation between broader spectra and larger vorticity, rather than a mechanistic link. Similarly, the enhancement of NIW kinetic energy in regions of strong vorticity may influence the global average, although comparisons between strong cyclonic and anticyclonic vorticity remain robust. More generally, interpreting global averages requires care, as they conflate diverse dynamical regimes.

Furthermore, we note that wind forcing typically excites a range of vertical modes \parencite[e.g.,][]{gill1984,thomas2024near}, each with different degrees of wave dispersiveness (Fig.~\ref{fig:eps}). The drifters feel the combination of these modes and may exhibit, even in a given region, behaviour that is a mix of strongly dispersive low modes and weakly dispersive higher modes. The spectra estimated from the drifters may reflect that mix. The strength of the imprint of strongly vs.\ weakly dispersive waves depends on their relative amplitudes, and a given diagnostic may be more sensitive to strongly or weakly dispersive waves, requiring nuance in the interpretation.

\begin{figure}
  \centering
  \includegraphics{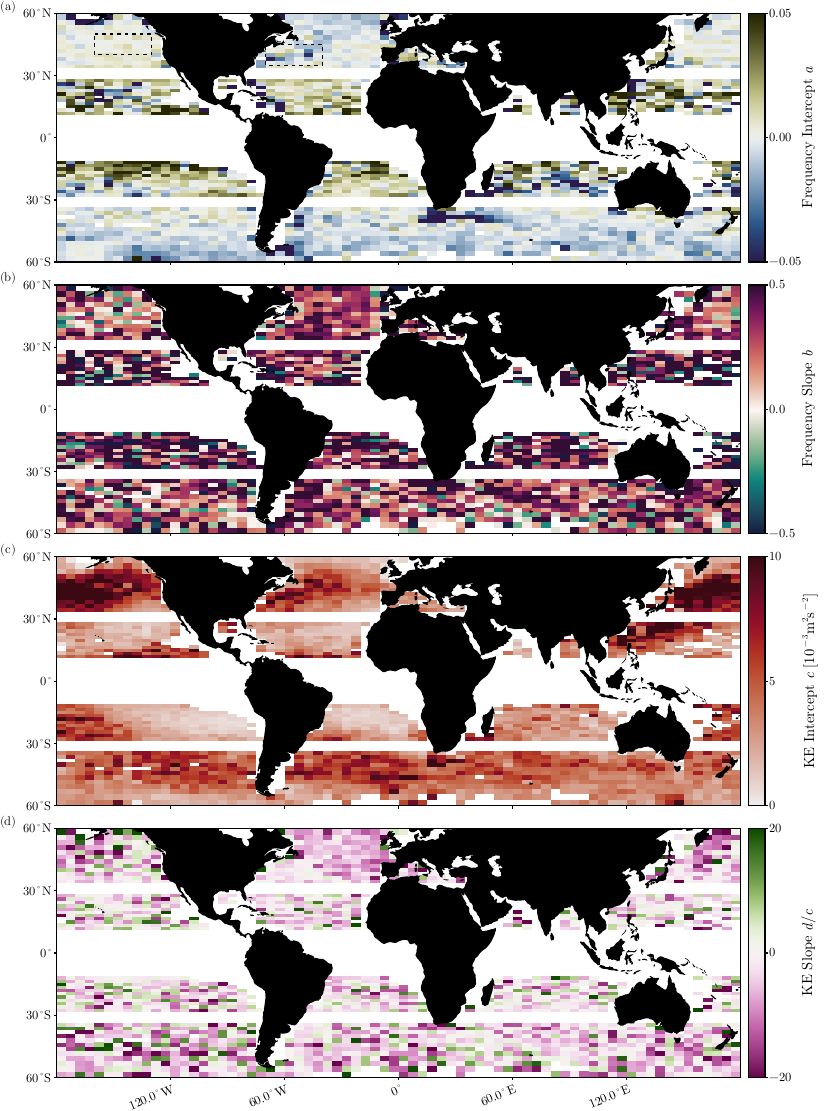}
  \caption{Geography of the vorticity dependence of NIW spectral properties. (a)~Intercept and (b)~slope of the linear fit to the NIW frequency shift~$\omega_I/f$ as a function of vorticity~$\zeta$. The black boxes in~(a) outline the two regions to be investigated further in Fig.~\ref{fig:regional}. (c)~Intercept and (d)~slope of the linear fit to the NIW kinetic energy~$K$ as a function of vorticity~$\zeta$. Note that we show the slope~$d$ normalized by the intercept~$c$ because otherwise it primarily reflects the patterns seen in panel~(c). 95\% confidence intervals are shown in Suppl.~Fig.~2.}
  \label{fig:ab}
\end{figure}

With these cautionary notes in mind, we calculate linear fits for $\omega_I$ and $K$ across the globe. To improve the statistics, we compute $\omega_I(\zeta)$ within each spatial bin and then average these curves to obtain a uniform \ang{5}$\times$\ang{5} discretization. A linear fit is then applied to the mean $\omega_I(\zeta)$ curve within each \ang{5}$\times$\ang{5} bin. The resulting intercept $a$ exhibits significant spatial structure (Fig.~\ref{fig:ab}a). Because of the lack of a  strong preference for cyclonic vs.\ anticyclonic vorticity at the mesoscale \parencite{chelton_global_2011}, the intercept~$a$ can be interpreted as the mean frequency shift with any bias in drifter locations toward cyclones removed. Notably, the spatial structure of $a$ (Fig.~\ref{fig:ab}a) closely resembles that of the mean $\omega_I$ (Fig.~\ref{fig:global_spec_prop}a), suggesting that these patterns are not an artifact of sampling bias. Strong negative intercepts are observed in regions of high eddy kinetic energy, such as western boundary currents and the Antarctic Circumpolar Current. As shown in~\eqref{eq:sd_freq_shift}, strongly dispersive NIWs are expected to exhibit negative frequency shifts proportional to mesoscale kinetic energy. These negative intercepts may therefore signal the excitation of strongly dispersive NIWs. Positive frequency shifts, on the other hand, are likely due to equatorward propagation of NIWs, although we show below that weakly dispersive waves interacting with mesoscale eddies are also expected to produce positive shifts.

\begin{figure}
    \centering
    \includegraphics{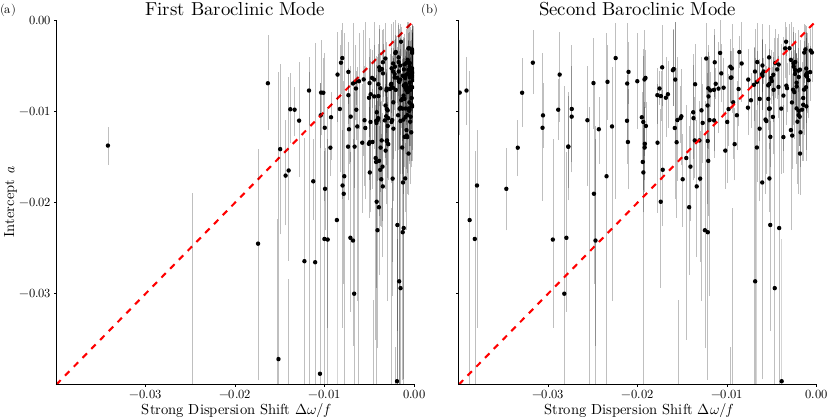}
    \caption{Comparison to the strong-dispersion prediction~\eqref{eq:sd_freq_shift}. Scatter plot of the intercept~$a$ of the frequency--vorticity fit in each spatial bin vs.\ the predicted frequency shift in the strong-dispersion limit for (a)~the first and (b)~the second baroclinic modes. The 1:1 line is shown in red.}
    \label{fig:SD_scatter}
\end{figure}

Negative frequency shifts associated with strongly dispersive waves may explain the negative values of the intercept $a$ observed in western boundary currents in the drifter data. A fraction of the near-inertial wind work excites strongly dispersive low modes \parencite[e.g.,][]{thomas2024near}, which are associated with a negative frequency shift proportional to the local mesoscale kinetic energy. Whether the full NIW signal also exhibits a negative frequency shift depends on the extent to which the wind excites these modes and the magnitude of their individual frequency shifts. The close spatial alignment between regions of high mesoscale kinetic energy and negative $a$ values leads us to hypothesize that large mesoscale kinetic energy is driving large negative frequency shifts in strongly dispersive waves, which are then expressed in the full signal. To test this idea, we compare the magnitude of the negative intercepts with the prediction from~\eqref{eq:sd_freq_shift} for strongly dispersive waves. To estimate this prediction from observations, we require both the mesoscale kinetic energy and the deformation radius. Using the streamfunction from altimetry, we calculate the mesoscale kinetic energy as $\frac{1}{2}|\nabla\psi|^2$. The deformation radius is obtained by solving the baroclinic eigenvalue equation \parencite{smith2007geography} using climatological data from the Estimating the Circulation and Climate of the Ocean (ECCO) state estimate version 4 release 4 \parencite{fukumori2020synopsis,forget2015ecco}. We perform this calculation for the first two baroclinic modes, which are generally strongly dispersive (Fig.~\ref{fig:eps}). The measured~$a$ from the drifter data reflects a mixture of frequency shifts from all baroclinic modes, as well as possible contributions from the $\beta$-effect, so a one-to-one correspondence is not expected. Nonetheless, we can assess whether the magnitude of the predicted shifts for strongly dispersive waves are consistent with the observed~$a$. The observed~$a$ generally lies between the values predicted for the first and second baroclinic modes (Fig.\ref{fig:SD_scatter}). It is therefore plausible that the negative values of~$a$ result from the presence of strongly dispersive waves. 

The slope $b$, while somewhat noisy, exhibits little spatial variability (Fig.~\ref{fig:ab}b). With few exceptions, its value remains close to the global mean. The drifter data thus provides evidence that mesoscale eddies modulate NIW spectra across most of the ocean. The North Pacific stands out as the only region where the slope $b$ is consistently weaker. This region was the site of the Ocean Storms Experiment, whose results similarly indicated weak mesoscale modulation—consistent with our findings. In Section~\ref{sec:sims}, we show that, theoretically, the slope is expected to asymptote to a constant value in the weak-dispersion regime, which helps explains the limited spatial structure in $b$.

The intercept $c$ (Fig.~\ref{fig:ab}c) closely mirrors the spatial distribution of NIW kinetic energy in the ocean (cf., Fig.~\ref{fig:global_spec_prop}c). The slope $d$, while somewhat noisy, is predominantly negative (Fig.~\ref{fig:ab}d), indicating that the concentration of NIWs into anticyclones is a widespread feature of the global ocean. That said, there are notable regions where the slope $d$ weakens. In particular, the Gulf Stream region exhibits a very modestly negative slope, indicating weakened concentration into anticyclones. In regions with strong eddies, which have relatively weaker wave dispersion (Fig.~\ref{fig:eps}), the tendency of waves to concentrate in anticyclones may be disrupted by advective straining \parencite{rocha2018}. While this map shows that anticyclonic concentration is a common aspect of NIW dynamics, it also highlights that the effect depends on wave dispersiveness in complex ways.

\begin{figure}
  \centering
  \includegraphics{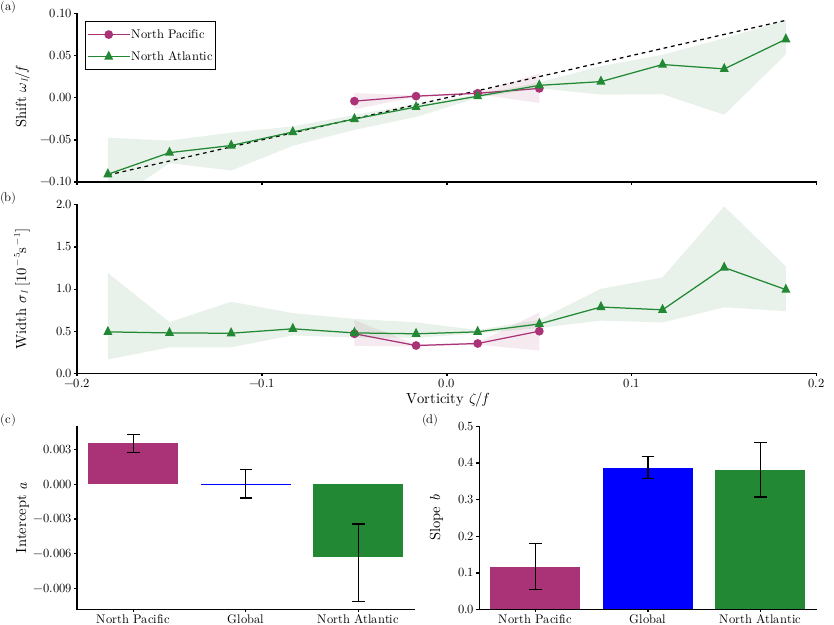}
  \caption{Regional vorticity dependence of the NIW spectral properties. (a)~NIW frequency shift~$\omega_I/f$ as a function of vorticity $\zeta$ when averaged over all spectra in the North Pacific (purple) and the North Atlantic (green) regions marked in Fig.~\ref{fig:ab}a. (b)~Same but for the NIW spectral width~$\sigma_I/f$. (c)~Intercept of the linear fits in~(a) and the global average (blue) for comparison. (d)~Same but for the slope of the linear fit.}
  \label{fig:regional}
\end{figure}

As the slope $b$ is somewhat noisy, and to better understand the behavior of the spectral width, we average the spectra over two representative regional domains (shown as black boxes in Fig.~\ref{fig:ab}b). These two regions differ markedly in the characteristics of their mesoscale eddy fields. In the North Pacific box, eddies are weak, while the North Atlantic box encompasses the energetic Gulf Stream rings. In the North Pacific, we observe smaller slopes and a positive frequency shift; in the North Atlantic, we find steeper slopes and a negative frequency shift. The regional averages support this distinction (Fig.~\ref{fig:regional}a). The spectral width is greater in the North Atlantic than in the North Pacific (Fig.~\ref{fig:regional}b). The weak slope observed in the North Pacific appears to be uncommon across the global ocean, with the possible exception of a few regions in the Southern Ocean, although the statistical uncertainty is higher there. The weak eddy field in the North Pacific corresponds to higher wave dispersiveness overall, favoring strongly dispersive waves and reducing mesoscale modulation. In the North Atlantic, we see the negative frequency shift indicative of strongly dispersive waves, but also clear modulation of the frequency shift by mesoscale vorticity, consistent with weakly dispersive wave behavior. Once again, we observe a complex picture in which the winds excite multiple vertical modes, each with a different degree of dispersiveness. The spectral characteristics reflect the aggregate influence of all these modes.

We have identified several characteristic properties of NIW spectra on a global scale that appear linked to interactions with mesoscale eddies. Regions of negative frequency shift are collocated with energetic western boundary currents and the ACC. Additionally, the slope of the frequency shift with vorticity shows remarkably little variation across the ocean, typically remaining close to the global average of $\sim$\num{0.4}. In a few localized regions, the slope flattens to near zero. Finally, the spectral width is elevated in energetic regions. To investigate the physical origins of these patterns, we examine the spectral properties of NIWs in an idealized model of NIW–mesoscale interactions.

\section{Idealized Simulations of NIWs}
\label{sec:sims}

Simulations of the YBJ equation provide an idealized framework for investigating the impact of mesoscale interactions on NIW spectral properties across different regimes, and for assessing the extent to which NIW–mesoscale interactions shape the observed NIW spectra from drifters. \textcite[][hereafter XV]{xie2015} proposed a model that couples the YBJ equation to a quasi-geostrophic (QG) flow. We simulate the XV model across a range of wave dispersiveness values. Using these simulations, we advect Lagrangian particles through the NIW field to generate synthetic drifter observations. These synthetic trajectories can then be used to calculate NIW spectral properties under controlled conditions.

Following \textcite{rocha2018}, the XV model for a given vertical mode is:
\begin{gather}
  \pp{\phi}{t} + \J(\psi,\phi) + \frac{i\zeta}{2}\phi - \frac{if\lambda^2}{2}\nabla^2\phi = -\nu_\phi\nabla^4\phi,\\
  \pp{q}{t} + \J(\psi,q) = -\kappa_q\nabla^4 q,\\
  q = \nabla^2\psi + q_w, \quad q_w = \frac{1}{f_0}\left[\frac{1}{4}\nabla^2|\phi|^2 + \frac{i}{2} \J(\phi^*,\phi)\right],
\end{gather}
where $q$ is the potential vorticity (PV), $\nabla^4 = \nabla^2\nabla^2$ is the biharmonic operator, $\nu_\phi$ is the hyper-diffusivity for NIWs, and $\kappa_q$ is the hyper-diffusivity for the QG flow. The biharmonic terms are included for numerical stability. This formulation represents a 2D version of the XV model in which the mesoscale flow is assumed to be barotropic, and the NIW velocity has been expanded in vertical baroclinic modes. These equations are solved on a doubly periodic domain using the Dedalus pseudospectral solver \parencite{burns2020}. Simulation parameters are provided in Appendix~B.

\begin{figure}
  \centering
  \includegraphics{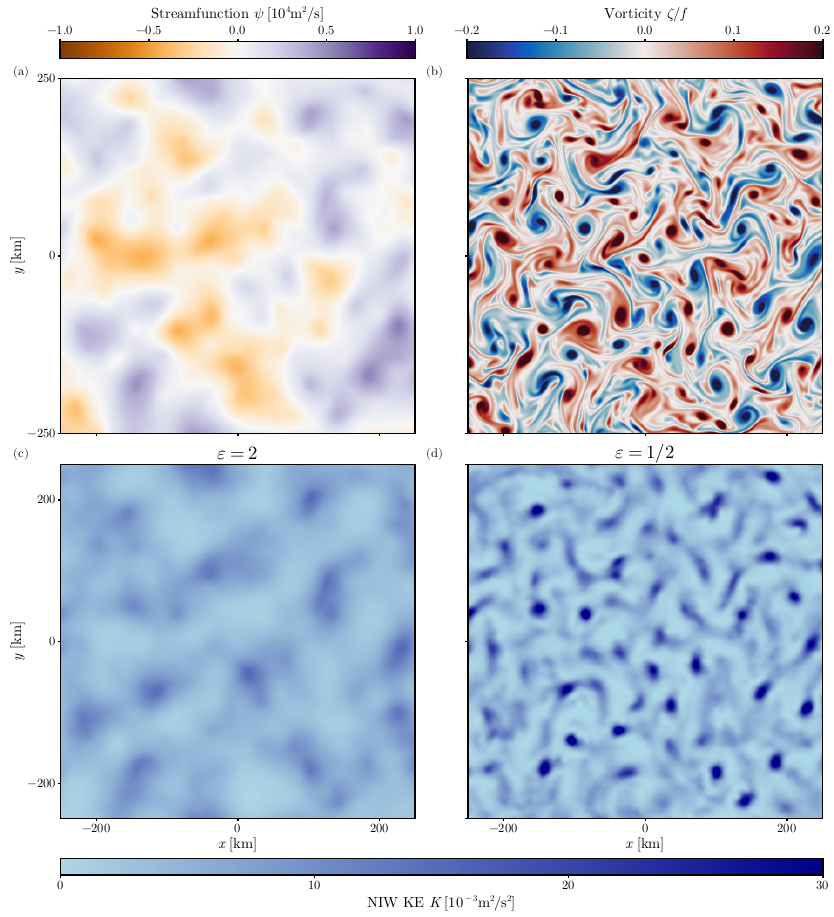}
  \caption{Idealized XV simulations of NIW--mesoscale interaction. (a)~Snapshot of the streamfunction~$\psi$ in the simulations with~$\varepsilon=2$ at~$t = \qty{4}{\days}$. (b)~Snapshot of the corresponding vorticity. (c)~Snapshot of the corresponding NIW kinetic energy density. Note that the structure resembles the streamfunction in panel~(a). (d)~Same but for~$\varepsilon=1/2$. Note that there is more small-scale structure that resembles the vorticity field. (At this early stage, the vorticity looks the same for $\varepsilon=2$ and~$1/2$. The impact of $q_w$ on the QG flow has not yet had a chance to cause the two simulations to drift apart. As such, we only show the vorticity for $\varepsilon=2$.)}
  \label{fig:wave_snapshots}
\end{figure}

We begin by evolving the QG flow for \qty{25}{\days} without waves (i.e., $\phi = 0$). After this spin-up period, we introduce a horizontally uniform NIW field with magnitude $U_w$. The waves are then allowed to evolve for an additional \qty{60}{\days}. We repeat the simulations for various values of wave dispersiveness. The qualitative structure of the wave field varies significantly with dispersiveness. For $\varepsilon \gg 1$, the waves exhibit large-scale structure resembling the streamfunction (Fig.~\ref{fig:wave_snapshots}a). For $\varepsilon \ll 1$, in contrast, the waves take on smaller-scale features more akin to the vorticity (Fig.~\ref{fig:wave_snapshots}b).

For each experiment, we advect Lagrangian particles to mimic the drifters. Particles are seeded in the model domain with a spacing of \qty{25}{\kilo\meter} and advected forward and backward for \qty{10}{\days}, yielding a total 20-day trajectory per particle. New particles are reseeded every \qty{10}{\days}. We use the Ocean Parcels package \parencite{delandmeter2019parcels} to perform the Lagrangian advection. The advection velocity is derived from $\psi$, which represents the Lagrangian-mean streamfunction that transports tracers \parencite{wagner2015available}. Particles sample the NIW velocity $\phi e^{-ift}$ at hourly intervals.

\begin{figure}
  \centering
  \includegraphics{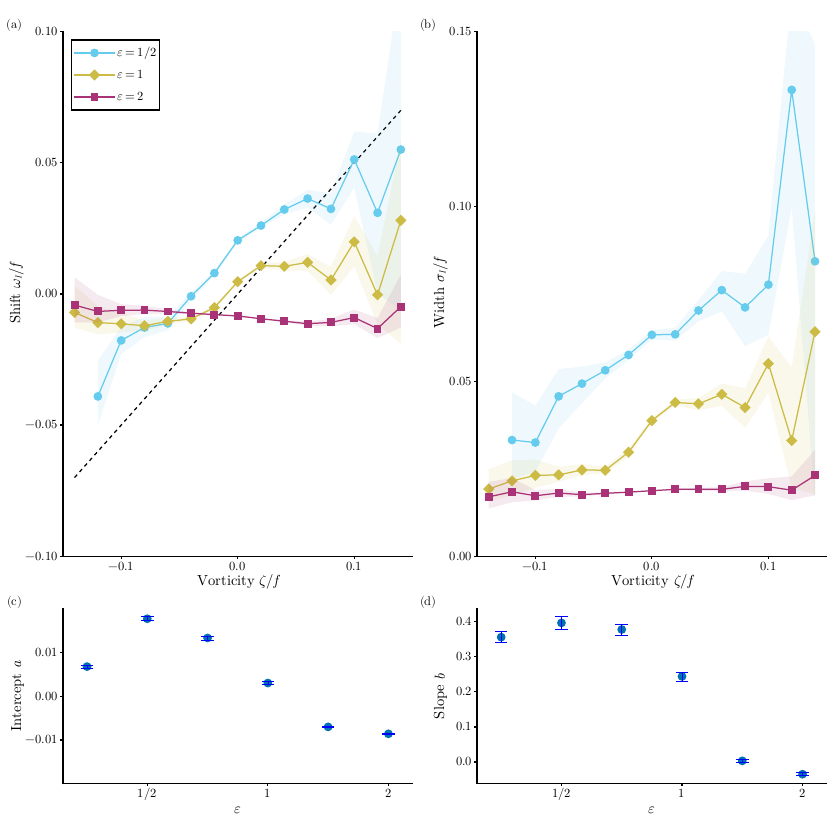}
  \caption{Spectral characteristics of the NIW peak in the XV simulations. (a)~NIW frequency shift~$\omega_I/f$ as a function of vorticity~$\zeta$ for $\varepsilon=1/2$ (weak dispersion, blue), $\varepsilon=1$ (transition case, yellow), and $\varepsilon=2$ (strong dispersion, purple). The $\zeta/2$ line is shown in black. (b)~Same but for the NIW spectral width $\sigma_I$. (c)~Intercept~$a$ of the linear fit as a function of $\varepsilon$. (d)~Same but for the slope~$b$.}
  \label{fig:YBJ_res}
\end{figure}

The dependence of the frequency shift and spectral width on vorticity varies markedly with $\varepsilon$. In the strong-dispersion regime ($\varepsilon \gg 1$), there is no modulation of either the shift or the width by vorticity. In contrast, in the weak-dispersion regime ($\varepsilon \ll 1$), both quantities show clear modulation with vorticity (Fig.~\ref{fig:YBJ_res}a,b): the frequency shift exhibits a positive slope with vorticity, while the width is enhanced in cyclonic regions. The intercept~$a$ is negative in the strong-dispersion regime and variable but generally positive in the weak-dispersion regime (Fig.~\ref{fig:YBJ_res}c). The slope approaches zero in the strong-dispersion limit and asymptotes to a value of $\sim$\num{0.4} in the weak-dispersion limit (Fig.~\ref{fig:YBJ_res}d).

These XV simulations produce spectra that exhibit many of the salient features observed in the NIW spectra from drifters, supporting the idea that these features arise from NIW–mesoscale interactions. While idealized, however, the simulations are still complicated, making it difficult to isolate the physics responsible for the imprint of mesoscale eddies on NIW spectra. To better illuminate these dynamics, we turn to a much simpler example: NIWs interacting with a single vortex dipole.

\section{NIWs in an Idealized Vortex Dipole}

\begin{figure}
  \centering
  \includegraphics{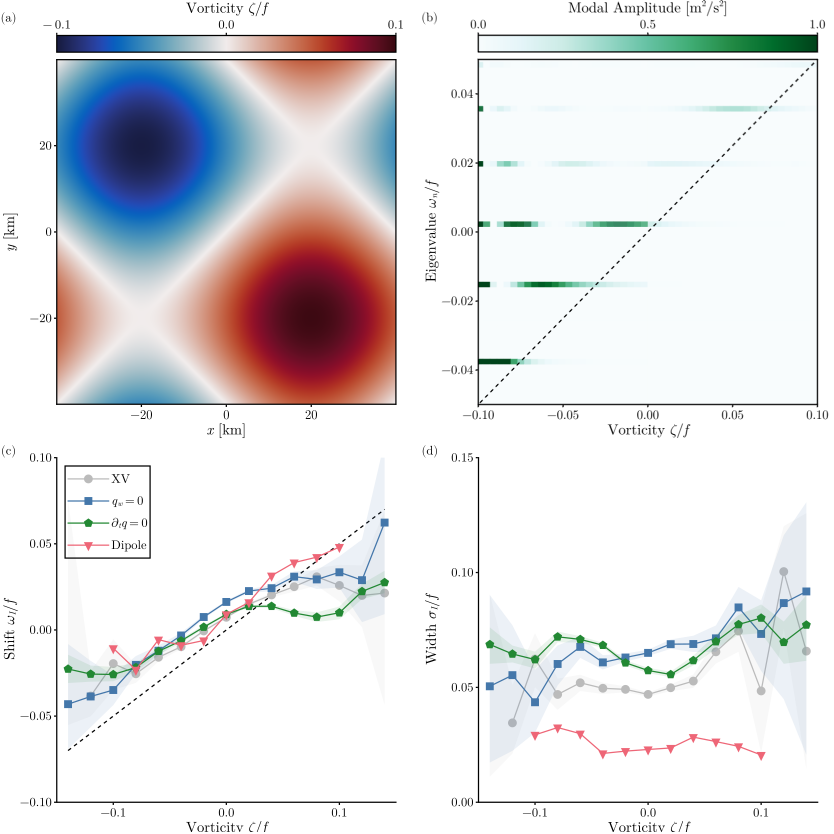}
  \caption{NIW spectral characteristics for progressively idealized simulations. (a)~Vorticity of the dipole flow. (b)~Modal amplitude as a function of vorticity for each eigenmode of the dipole flow. (c)~NIW frequency shift~$\omega_I$ as a function of~$\zeta$ for the simulations with the XV coupling turned off (blue), with the streamfunction evolution turned off (green), and for the dipole flow (pink). The $\zeta/2$ line is shown in black. (d)~Same but for the NIW spectral width~$\sigma_I$. Note that we have re-dimensionalized the dipole flow with a streamfunction magnitude $\Psi$ that matches the XV simulations and a domain length of \qty{80}{\kilo\meter}. Furthermore, the dipole lines come from YBJ simulations with the dipole as the background field. We also calculated these lines from the eigenmodes described above and found that the two methods agree.}
  \label{fig:dipole}
\end{figure}

The goal of this section is to understand the physics of NIW–mesoscale interactions and their effect on NIW spectra in the simplest possible context. \textcite{conn2025regimes} adopted a spectral approach to solving the YBJ equation, calculating the eigenvalues and eigenvectors of the YBJ operator---a useful method for studying NIW spectra. We begin by non-dimensionalising the YBJ equation and rewriting it as
\begin{equation}
  i\frac{\partial\tilde{\phi}}{\partial \tilde{t}} = H\tilde{\phi} = - \frac{\varepsilon^2}{2} \nabla^2 \tilde{\phi} - i \J(\tilde{\psi}, \tilde{\phi}) + \frac{\tilde{\zeta}}{2}\tilde{\phi},
\end{equation}
where $H$ is the YBJ operator, the tildes indicate non-dimensionalized fields, and the operators are understood to take derivatives with respect to the non-dimensional variables. We then consider the spectrum of the operator $H$ by solving for the eigenvalues and eigenfunctions of~$H$,
\begin{equation}
    H\tilde{\phi}_n = \tilde{\omega}_n\tilde{\phi}_n,
\end{equation}
where $\tilde{\phi}_n$ is an eigenfunction, $\tilde{\omega}_n$ is the associated eigenvalue, and $n$ is a label for the eigenfunction. Throughout this section, we consider a simple vortex dipole \parencite[Fig.~\ref{fig:dipole}a][]{asselin2020refraction,conn2025regimes},
\begin{equation}
    \tilde{\psi} = \frac{1}{2}\left(\sin \tilde{x} - \sin \tilde{y}\right),
\end{equation}
consisting of a cyclone and anticyclone on a doubly period domain with side length~$2\pi$.

In the strong-dispersion limit~$\varepsilon \gg 1$, the nearly uniform mode dominates and results in a strongly peaked spectrum at a weakly shifted frequency as described by~\eqref{eq:sd_freq_shift}. Therefore, we here focus our attention on the weak-dispersion limit. For the moment, let us ignore advection and write the eigenvalue equation as
\begin{equation}
    -\frac{\varepsilon^2}{2}\tilde{\nabla}^2\tilde{\phi}_n = \left(\tilde{\omega}_n-\frac{\tilde{\zeta}}{2}\right)\tilde{\phi}_n.
\end{equation}
From this equation, one can see that $\tilde \zeta$ acts like a potential in the Schr\"odinger equation, and for a given mode with eigenvalue $\tilde{\omega}_n$, the eigenfunction is oscillatory where $\tilde{\omega}_n > \tilde{\zeta}/2$ and evanescent where $\tilde{\omega}_n < \tilde{\zeta}/2$. In other words, lower-frequency modes are screened out in regions of higher vorticity. At a given vorticity~$\tilde \zeta$, one should thus see only modes that have a frequency of~$\tilde \zeta/2$ or greater. While a number of modes contribute to the NIW frequency spectrum at low vorticities, a large-scale wind forcing, here modeled as a uniform forcing, projects most strongly onto the low modes. As a result, the NIW frequency increases monotonically with~$\tilde \zeta$, with a slope of the frequency--vorticity curve close to~$1/2$. This argument is admittedly hand-wavy and neglects advection, but the more rigorous treatment in \textcite{conn2025regimes} supports this intuition.

To give a concrete example---now including advection---we compute the eigenfunctions and eigenvalues for the dipole flow. The numerical procedure is described in \textcite{conn2025regimes}. We also compute the projection coefficient of a uniform initial condition onto each mode, and we define the modal amplitude as the mean, taken along contours of constant vorticity, of the modulus squared of each eigenmode multiplied by its projection coefficient. The results clearly illustrate the screening of modes with~$\tilde{\omega}_n$ less than the local $\tilde{\zeta}/2$ (Fig.~\ref{fig:dipole}b). The strong projection onto the lowest eigenmodes ensures that the mean frequency remains close to the $\tilde{\zeta}/2$ line, even in the core of the anticyclone (Fig.~\ref{fig:dipole}c).

The dipole calculation therefore provides a physical explanation for the observed modulation of the frequency shift by vorticity. This modulation occurs across much of the ocean (Fig.~\ref{fig:ab}b), implying the near-universal presence of weakly dispersive wave modes that are refracted by mesoscale vorticity. The dipole calculation also predicts, however, that the spectral width should be elevated in anticyclones (Fig.~\ref{fig:dipole}b,d), whereas the simulations and, to a lesser extent, the observations show the opposite. To investigate this discrepancy, we return to the XV model from Section~\ref{sec:sims}, running it with varying levels of complexity.

First, we disable the influence of NIWs on the mesoscale by setting $q_w = 0$ in the XV equations. The QG flow continues to evolve, but there is no feedback from the waves. Next, we run a simulation in which we turn off the time evolution of the QG flow altogether: the initial condition is the same as before, but the flow is held fixed in time after the spin-up. Finally, we perform a simulation in which we replace the initial condition with the dipole flow (re-dimensionalized to have $\Ro = 0.1$). This is a stationary solution to the QG equations and therefore does not evolve in time. These three simulations form a hierarchy of complexity, ranging from the simplest dipole case to the full XV model.

Setting $q_w = 0$ has minimal impact on the results compared to the full XV simulation, leading only to a slight increase in the spectral width (Fig.~\ref{fig:dipole}c,d). The $\partial_t q$ simulation also leaves the frequency--vorticity relation largely unchanged, except that there is a drop in the frequency shift at positive~$\zeta$ (Fig.~\ref{fig:dipole}c). It shows similar spectral width as the XV and $q_w = 0$ simulations. The dipole simulation recovers a frequency--vorticity relation with a slope close to~$1/2$, but it has a markedly decreased spectral width and no increase with~$\zeta$ (Fig.~\ref{fig:dipole}d). Instead, as expected from the modal picture, the spectral width is greatest at negative~$\zeta$.

It appears that the key complexity required to produce spectra that qualitatively resemble the drifter observations is the presence of a sea of eddies of varying amplitude. Even when the eddies are stationary---just one step up in complexity from the dipole simulations---we observe behavior in both the frequency shift and the spectral width that resembles the observations. In the dipole flow, averaging over vorticity amounts to averaging within a single eddy, whereas in the turbulent QG flow, the averaging spans multiple eddies with varying structures. In a sea of eddies, a given cyclone may have stronger cyclones nearby, meaning it is no longer necessarily at the top of the $\zeta/2$~landscape that acts like a potential. Eigenmodes with frequencies higher than that of the local cyclone can be excited and felt within it, thereby increasing the spectral width. The same does not hold for anticyclones, because modes with energy lower than the local $\zeta/2$ are screened. The result is an asymmetry in spectral width between cyclones and anticyclones as observed in the simulations. The behavior of the spectral width, therefore, is not governed solely by NIWs interacting with isolated eddies, but instead emerges from interactions within a sea of eddies.

\section{Discussion}
\label{sec:discussion}

The drifter observations, interpreted in the context of NIW–mesoscale interactions, reveal a complex picture of NIW behavior. Spatial variability in the mesoscale flow is clearly imprinted onto the NIW spectral properties. Even within a single region, however, we observe signatures of both weakly and strongly dispersive waves. This arises from the wind forcing exciting multiple vertical modes, each characterized by a different degree of wave dispersiveness. The drifters sample a combination of these modes and their associated behaviors. While it is beyond the scope of this manuscript to determine which vertical modes are excited in each region and how dispersive they are, this remains a key factor in the evolution of NIWs. \textcite{thomas2024near} performed such a projection in regions of the North Atlantic and North Pacific, and \textcite{conn2025regimes} calculated climatological wave dispersiveness for the first four baroclinic modes (see Fig.~\ref{fig:eps}). Extending \citeauthor{thomas2024near}'s calculations to a global scale would provide valuable insight into how the structure of wind forcing and background stratification influence the spectral characteristics of NIWs.

We found negative shifts in the NIW frequency in regions of high mesoscale energy (Fig.~\ref{fig:global_spec_prop}a) and showed that these are plausibly associated with strongly dispersive NIWs. We found global-scale positive frequency shifts equatorward of \ang{30} latitude (Fig.~\ref{fig:global_spec_prop}a). \textcite{elipot2010modification} attributed this feature of the observations to the $\beta$-effect, which causes equatorward propagation of NIWs and therefore produces a positive shift. We also showed that weakly dispersive NIWs interacting with a background mesoscale eddy field are expected to produce a positive shift (Figs.~\ref{fig:YBJ_res},~\ref{fig:dipole}). The global pattern and the increasingly positive shifts toward the equator, where $\beta$ is larger, support the equatorward propagation mechanism. \textcite{garrett2001near} calculated the meridional distance that NIWs propagate in the time it takes them to reflect off the seafloor and return to the surface, finding $\sim$\qty{400}{\kilo\meter} except for very close to the equator and pole. The associated frequency shift, normalized by~$f$, increases markedly toward the equator and is of order $0.1$. The observed frequency shift is somewhat smaller but has the right order of magnitude and pattern. The distance is not large enough for NIWs generated under mid-latitude storm tracks to contribute substantially to subtropical NIW energy, but it is plausible that the subtropical NIW peak consists of a mix of NIWs generated locally and up to a few degrees poleward \parencite[cf., e.g.,][]{raja2022near}. The origin of these frequency shifts should be further investigated using realistic simulations or in-situ field campaigns, which are better suited than surface drifters to distinguish the sources of NIW energy.

In this study, we have considered time means of all spectral quantities and have therefore ignored any potential seasonality in the NIW spectral properties. Seasonal variations in mesoscale eddies and the deformation radius are generally weak over much of the ocean \parencite[cf., e.g.,][]{chelton1998geographical} and so these are unlikely to significantly alter these results. However, both the wind forcing and stratification exhibit substantial seasonal variability \parencite[cf., e.g.,][]{alford2007seasonal,sallee2021summertime}. This seasonal variability could impact the projection of the wind forcing onto the different modes, changing the overall properties of the NIW spectra. A global-scale analysis of the seasonality of NIW spectral characteristics—similar to the approach taken in this paper—would likely be difficult at present due to the limited number of observations across much of the ocean. Regional analyses may prove feasible.

Strictly, the decomposition of NIWs into vertical modes is only valid under the assumption of a barotropic background flow. This assumption is clearly violated in the ocean, where mesoscale eddies often exhibit significant vertical structure. Interestingly, however, accounting for baroclinicity in the mean flow does not appear necessary to explain the spectral characteristics observed by the drifters, at least qualitatively. More work is needed to understand the spectral properties expected for NIWs interacting with a baroclinic mean flow.

In the real ocean, vorticity is a multiscale field, which we characterize using the heavily smoothed altimetry product. This vorticity field is further averaged along drifter trajectories. One might worry that substantial submesoscale vorticity is being missed and could be influencing the NIWs. We argue that this may be so, but dispersion acts as a natural filter for vorticity. In spectral space, dispersion scales like~$k^2$, where $k$ is the wavenumber. Thus, while vorticity tends to be stronger at smaller scales, so is dispersion. There must exist a scale below which dispersion is sufficient to balance refraction. The fact that we observe such large slopes in the frequency shift as a function of the mesoscale~$\zeta$ suggests that this relevant scale is by and large resolved by the altimetry. Indeed, shortening the length of drifter trajectory chunks reduces the slope; for example, using \qty{6}{\day} chunks lowers the slope to $\sim$\qty{0.25} (not shown). This implies that, on average, NIWs are not responding to local, small-scale vorticity features.

The observed spectral characteristics of NIWs have implications for filtering NIWs from raw signals that contain multiple types of motion. \textcite{rama2022importance} recently proposed using an adaptive filter in which the effective frequency $f + \zeta/2$ serves as a lower bound on a high-pass filter to separate NIWs from other flows. Our results suggest that this approach may not always be appropriate. This expression for the lower frequency limit is derived from ray-tracing, which is valid only in the weakly dispersive limit. Our results show clear evidence of strongly dispersive waves as well, for which there is no comparable modulation of the NIW frequency. Applying such an adaptive filter would introduce a cyclonic–anticyclonic bias when extracting strongly dispersive NIWs. Moreover, even for weakly dispersive waves, the $\zeta/2$ scaling is not universal. The drifter observations typically exhibit a slope closer to $\sim$0.4, with non-negligible NIW variance below the $f + \zeta/2$ cut-off. Using this cut-off would result in severe underestimation of NIW energy in cyclonic regions compared to anticyclonic ones. Small-scale vorticity, as discussed above, further complicates this approach. An adaptive approach that first locates and characterizes the NIW peak seems more promising.

Finally, we note that these results have implications for NIW-induced mixing on a global scale. Numerous observational studies have shown that the concentration of NIWs into anticyclones can lead to enhanced NIW-induced mixing at depth \parencite[e.g.,][]{kawaguchi2016enhanced,martinez2019near,sanford2021stalling}. Our results indicate that this concentration is fairly ubiquitous across the ocean. Therefore, in regions of high NIW energy, the interaction between mesoscale eddies and NIWs may play a significant role in shaping patterns of upper-ocean mixing. Future work should aim to quantify the magnitude of this effect.

\section{Conclusions}

Using observations of near-inertial waves from the global array of drifters, we have mapped the geography of NIW–mesoscale interactions across the world ocean. Our results indicate that mesoscale eddies influence the spectral characteristics of NIWs nearly everywhere. We also found that the YBJ equation, which describes the evolution of NIWs in a background mesoscale eddy field, explains much of the observed phenomenology of NIW spectra. Additionally, we observed evidence of both weakly and strongly dispersive NIWs, suggesting that wind forcing excites vertical modes for which dispersion has varying importance.

To quantify the modulation of NIWs by mesoscale eddies, we considered a linear fit to the relationship between the NIW frequency shift and the vorticity. The slope of this fit is $\sim$0.4 in the global mean and takes a similar value across much of the ocean---except in the North Pacific and possibly some regions of the Southern Ocean, where the slopes are close to zero. Typical slopes are close to the 0.5 value predicted by ray-tracing, although the XV simulations showed that the slope should asymptote to a somewhat smaller constant value in the weak-dispersion limit, offering an explanation for the lack of spatial variability in the slope on a global scale. Additionally, the dipole flow demonstrated that this slope arises from the screening of eigenmodes of the YBJ equation in regions in which the eigenvalue is less than~$\zeta/2$.

In contrast to the slope, the intercept exhibited much greater spatial structure when mapped globally. In highly energetic regions such as the western boundary currents and the ACC, the intercept was negative, while elsewhere it was weakly positive. This close spatial collocation led us to consider strongly dispersive waves as a possible explanation for the negative frequency shifts. Theoretical predictions for the frequency shift in the strong-dispersion limit were found to be of similar magnitude as the observed shifts, acknowledging that a one-to-one correspondence is not expected. The pattern of positive shifts pointed to the $\beta$-effect as a likely source, although weakly dispersive waves are also expected to produce positive shifts.

The NIW spectral width also showed a striking correspondence with the highly energetic western boundary currents and the ACC, where the width was elevated. This can be understood by considering the eigenfunctions of the YBJ equation. In a stronger eddy field, the potential wells are deeper, allowing more modes to be excited and thereby increasing the spectral width. The width typically increases with vorticity, a behavior that is expected for a turbulent field of eddies of varying size and reproduced by idealized XV and YBJ simulations.

Finally, we also considered how NIW kinetic energy depends on vorticity. The global mean showed a clear concentration into anticyclones, while the global distribution of mean NIW kinetic energy reflected the patterns of large-scale forcing. Examining the slope of NIW kinetic energy versus vorticity across the globe, we found that anticyclonic concentration is typical throughout much of the ocean. However, a few notable regions---the Gulf Stream region, in particular---exhibited a weak dependence of kinetic energy on vorticity, despite being highly energetic. This suggests that while concentration into anticyclones is widespread, strong advection can disrupt this tendency.

\section*{Acknowledgments}
The authors gratefully acknowledge support from NASA under grants 80NSSC22K1445 and 80NSSC23K0345, 80NSSC24K1652, and 80NSSC20K1140.

\section*{Data Availability Statement}

This study used data collected and made freely available by the NOAA Global Drifter Program accessed from \url{https://erddap.aoml.noaa.gov/gdp/erddap/index.html}. The code to run the XV simulations can be found at \url{https://github.com/scott-conn/2DYBJ-QG}. The SSH data is available from the E. U.’s Copernicus Marine Service at \url{https://doi.org/10.48670/moi-00148}.  The ERA5 reanalysis data is available from the Copernicus Climate Change Service (C3S) Climate Data Store at \url{https://doi.org/0.24381/cds.adbb2d47}. The ECCO density data is available from \url{https://doi.org/10.5067/ECG5D-ODE44}.

\appendix
\section{Least-Squares Fit}
Given some spectral estimate $\hat{S}$, we wish to find the parameters in Equation~\ref{eq:spectral_model} that minimize the residuals between $\hat{S}$ and the model. Writing the model as $S(\omega,p)$, where $p$ represents the set of model parameters, the residuals are defined as
\begin{equation}
    R(p) = \sum_n\left[\hat{S}_n - S(\omega_n,p)\right]^2,
\end{equation}
where $\omega_n$ are the discrete frequencies resolved by the data. We use the Levenberg–Marquardt algorithm to find the set of parameters $p^*$ which minimizes $R(p)$.

\section{Simulation Parameters}

\begin{table}
  \centering
  \begin{tabular}{ll}
    \toprule
    Parameter    &  Value\\
    \midrule
    Domain length $L$    & $4\pi\times10^5$\qty{}{\meter} \\
    Number of modes $N$ & 1024\\
    Coriolis frequency $f$ & \qty{1e-4}{\per\second}\\
    NIW speed $U_w$ & \qty{0.1}{\meter\per\second}\\
    Eddy speed $U_e$ & \qty{0.05}{\meter\per\second}\\
    Eddy wavenumber $k_e$ & $16\pi\times10^{-6}$\qty{}{\per\meter}\\
    Biharmonic viscosity $\nu_w$ & \qty{5e6}{\meter\tothe{4}\per\second}\\
    Biharmonic diffusivity $\kappa_e$ & \qty{5e6}{\meter\tothe{4}\per\second}\\
    Timestep $\Delta t$ & \qty{25}{\second}\\
    Simulation run time $T$ & \qty{5.184e6}{\second}\\
    \bottomrule
  \end{tabular}
  \caption{Parameters for the XV simulation}
  \label{tab:param}
\end{table}

For most of the simulation parameters, we follow \textcite{rocha2018}. These can be found in Table~\ref{tab:param}. We also follow \textcite{rocha2018} in initializing the QG flow by specifying the Fourier transform of~$\psi$ as
\begin{equation}
    \hat{\psi}(\vec{k}) = \frac{C}{\sqrt{|\vec{k}|\left[1+\left(\frac{|\vec{k}|}{k_e}\right)^4\right]}},
\end{equation}
where the hat indicates the Fourier transform, $|\vec{k}|$~is the magnitude of the wavevector, $k_e$~is an eddy wavenumber, and $C$ is a normalization constant. The value of~$C$ is chosen such that the domain-average kinetic energy density of the simulation is~$U_e^2 / 2$.

The value of~$\varepsilon$ is varied by changing the deformation radius~$\lambda$. We choose~$\lambda$ based on the RMS value of the streamfunction and $f$ to give the desired value of~$\varepsilon$ according to~$\varepsilon^2=f\lambda^2/\Psi$. 

\end{document}


\section{Confidence Intervals}
The spectral estimates, generated from multiple spectrograms, follow a $\chi^2$ distribution. The fitted parameters are a complicated non-linear function of the spectral estimate and so do we do not have an analytical form for their probability distributions. However, we can draw random samples from the distribution of the spectral estimate, put these through the fitting procedure, and build empirical distribution functions for the fitted parameters. Here we show the 95\% confidence interval half widths for the globally mapped parameters.

When using all spectrograms in a given box, the confidence intervals are generally quite narrow (Suppl.~Fig.~1). The exception is near the equator and in coastal regions where the confidence intervals can become large.

When binning the spectrograms by vorticity too, we necessarily have less spectrograms being averaged to get a spectral estimate. As a result, the confidence intervals are wider for properties that depend on the vorticity (Suppl.~Fig.~2). In general, the intercepts have a relatively narrow confidence interval, but the interval can be come large for the slopes. These plots reflect the distribution of drifter numbers, with the statistics being much better in the Atlantic where we have more drifter observations. For the frequency slope, the values are quite uncertain equatorward of \ang{30}, and in much of the Southern Ocean. The same is generally true for the KE slope.

\begin{figure}
    \centering
    \includegraphics{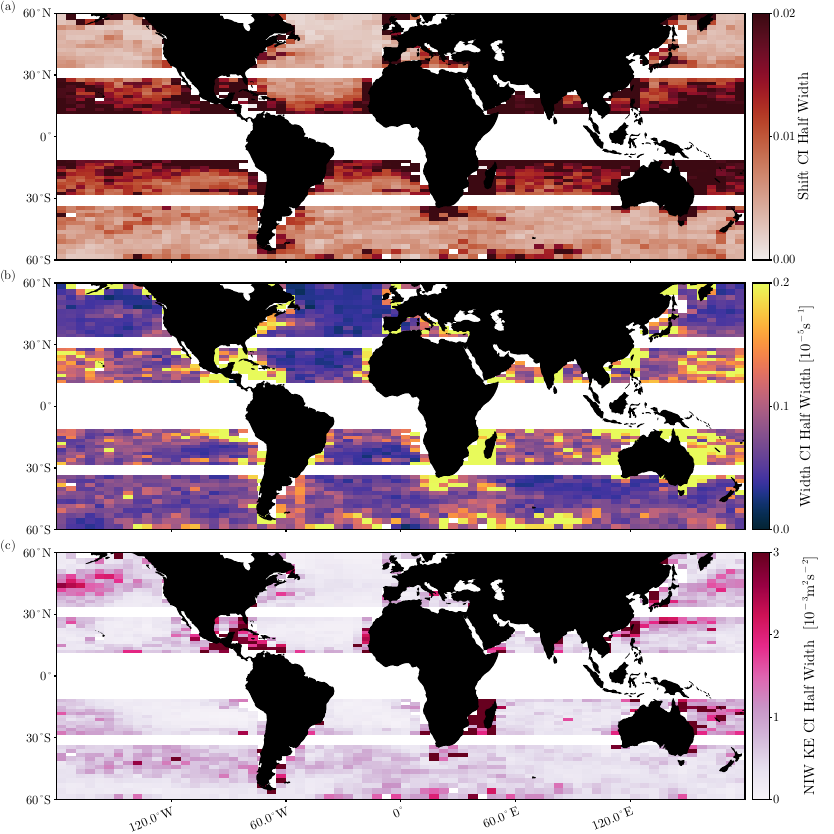}
    \caption{95\% confidence interval half width for the characteristics of the NIW peak: (a)~the frequency shift~$\omega_I/f$, (b)~the spectral width~$\sigma_I$, and (c)~the NIW kinetic energy~$K$.}
    \label{fig:S1}
\end{figure}

\begin{figure}
    \centering
    \includegraphics{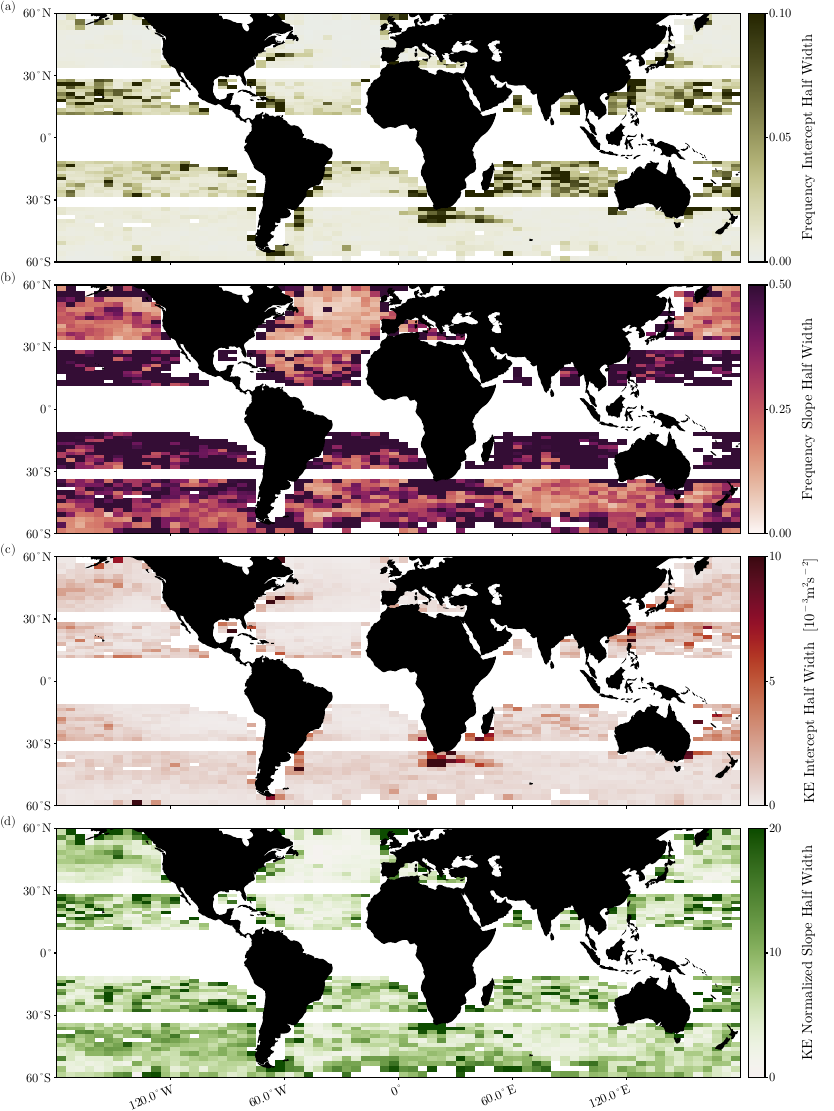}
    \caption{95\% confidence interval half width for the geography of the vorticity dependence of NIW spectral properties. (a)~Intercept and (b)~slope of the linear fit to the NIW frequency shift~$\omega_I/f$ as a function of vorticity~$\zeta$. (c)~Intercept and (d)~slope of the linear fit to the NIW kinetic energy~$K$ as a function of vorticity~$\zeta$. Note that we show the slope~$d$ normalized by the intercept~$c$ because otherwise it primarily reflects the patterns seen in panel~(c).}
    \label{fig:S2}
\end{figure}